\documentclass[pra,showpacs,twocolumn,superscriptaddress]{revtex4-1}
\usepackage{amsfonts}
\usepackage{amsmath}
\usepackage{amssymb}
\usepackage{graphicx}
\usepackage{epsfig}
\usepackage{bm}
\usepackage{enumerate}
\usepackage{multirow}

\newcommand{\ket}[1]{\left| #1 \right\rangle}

\newcommand{\bracket}[2]{\left\langle #1 | #2 \right\rangle}

\newcommand{\trace}[1]{\mathrm{Tr}\left( #1 \right)}

\makeatletter
\renewcommand*\env@matrix[1][\arraystretch]{%
  \edef\arraystretch{#1}%
  \hskip -\arraycolsep
  \let\@ifnextchar\new@ifnextchar
  \array{*\c@MaxMatrixCols c}}
\makeatother

\begin{document}


\title{Experimental scheme for qubit and qutrit symmetric informationally complete positive operator-valued measurements using multiport devices}




\author{Gelo Noel M. Tabia}
\email[Electronic address: ]{gtabia@perimeterinstitute.ca}
\affiliation{Perimeter Institute for Theoretical Physics, 31 Caroline Street North, Waterloo, Ontario N2L 2Y5, Canada}
\affiliation{Department of Physics and Astronomy and Institute for Quantum Computing, University of Waterloo,
200 University Avenue West, Waterloo, ON, N2L 3G1, Canada}

\begin{abstract}

It is crucial for various quantum information processing tasks that the state of a quantum system can be determined reliably and efficiently from general quantum measurements. One important class of measurements for this purpose is symmetric informationally complete positive operator-valued measurements (SIC-POVMs). SIC-POVMs have the advantage of providing an unbiased estimator for the quantum state with the minimal number of outcomes needed for full tomography. By virtue of Naimark's dilation theorem, any POVM can always be realized with a suitable coupling between the system and an auxiliary system and by performing a projective measurement on the joint system. In practice, finding the appropriate coupling is rather non-trivial. Here we propose an experimental design for directly implementing SIC-POVMs using multiport devices and path-encoded qubits and qutrits, the utility of which has recently been demonstrated by several experimental groups around the world. Furthermore, we describe how these multiports can be attained in practice with an integrated photonic system composed of nested linear optical elements.

\end{abstract}


\maketitle

\section{Introduction}

Quantum state tomography describes various methods for estimating the state of a quantum system from the statistics of a sufficiently informative set of measurements \cite{ivanovic1981, paris2004}. It continues to be one of the primary challenges in quantum information and quantum computing since typical quantum information processing tasks depend on the ability to determine the quantum state in a robust and efficient manner.

A general $d$-dimensional quantum state is characterized by $d^2-1$ real parameters, requiring a measurement with at least $d^2$ linearly independent outcomes. A measurement satisfying this requirement is said to be informationally complete \cite{prugovecki1977, busch1991}. A symmetric informationally complete positive operator-valued measurement (SIC-POVM) is a generalized measurement consisting of $d^2$ subnormalized projection operators with equal pairwise fidelity \cite{zauner1999, renes2004, scottgrassl2010}. In practical quantum state tomography, SIC-POVMs are important because they are known to be optimal measurements, according to fairly standard measures of efficiency: (i) they use the minimal number of expectation values, leading to less redundancy and faster convergence in state reconstruction, and (ii) they achieve the smallest error probability with respect to Hilbert-Schmidt distance between the estimated and ``true'' density operator for the measured system \cite{scott2006}.

The earliest known SIC-POVM experiment was performed on qubits by Durt, et al. \cite{durt2008} with a setup involving the simultaneous measurement of two sets of tetrahedral states that yielded the discrete Wigner distribution for two qubits. There were also experiments characterized as entanglement-assisted SIC-POVMs, where the photonic \cite{kurtsiefer2006} and NMR \cite{du2006} qubits were coupled to ancillary systems, allowed to interact, and then measured jointly. These experiments failed to measure the SIC-POVM operators directly and therefore, in a strict sense, are not true SIC-POVMs.

A practical approach for direct implementation of qutrit SIC-POVMs using linear optical elements was demonstrated by Medendorp et al. \cite{medendorpetal2010}, where the qutrit was encoded in polarizations states of a photon with two spatial modes. The photons propagated inside a storage loop containing a calcite beam displacer for performing the shift $X$ gate and three liquid crystal wave plates for performing phase $Z$ gates. The SIC-POVM projections were achieved with weak measurements, each of which involved diverting a tiny fraction of the input signal into a collection of half-wave plates and a partial polarizing beam splitter. A SIC-POVM measurement is done once the signal completes three rounds inside the storage loop. However, due to significant losses in the optical elements, the noisy weak projections differed quite significantly from ideal SIC-POVM projections. On the other hand, Schaeff et al. \cite{schaeffetal2012}, Meany et al. \cite{withford2012}, Spagnolo et al. \cite{spagnolo2012}, and Perruzo et al. \cite{obrien2011} have recently exhibited the promise of integrated optics for quality quantum information demonstrations. In this paper, we present an alternative experimental approach to realizing qubit and qutrit SIC-POVMs using waveguide-based multiport devices.

\section{Integrated optics and multiport devices}

The heart of quantum information processing using photons is a linear network of optical paths or modes designed to exhibit classical and nonclassical interference. In 2001, Knill, Laflamme, and Milburn demonstrated that linear optical devices supplemented by single photon sources and detectors are sufficient for realizing a scalable quantum computer \cite{klm2001}. In a standard optical implementation of a quantum circuit, photons propagate in free space and are manipulated using a combination of mirrors, phase shifters, and beam splitters. Due to their physical size, such bulk optical elements suffer from inherent phase instability and lack of scalability. Integrated optics technology offers a solution to these shortcomings by providing a waveguide-based architecture that achieves compact, efficient, and stable quantum optical performance. Physical realizations of integrated optics include silica-based planar lightwave circuit interferometers fabricated on a semiconductor chip \cite{kawachi1990, kato1998, obrien2008, matthews2009, smith2009, li2011, laing2010}, periodically poled lithium niobate waveguides in standard fiber optics technology \cite{prevedel2007, tanzilli2010, schaeff2011, bonneau2012, polster2012} and the direct-write femtosecond laser technique for manufacturing optical waveguides in dielectric materials \cite{hirao1996, nolte2003, obrien2009}, which allows one to realize three-dimensional photonic circuits, in contrast with the flat architecture of conventional lithography.

Moreover, integrated photonic circuits carry the practical advantage of realizing a multiport device, a blackbox for performing any discrete unitary operation through a sequence of qubit operations acting on various modes \cite{recketal1994}. More recent versions may employ multimode interference (MMI) devices in the integrated circuits \cite{obrien2011}, although at present the qudit operations these MMI devices can implement is still limited. An optical version of a multiport consists of integrated beam splitters and phase shifters with adjustable parameters that can be set to any $N$-dimensional unitary operator. Multiport devices provide a general and flexible platform for performing increasingly complex quantum optical circuits and may be used in various quantum optical experiments, such as the study of higher-dimensional entangled systems \cite{zukowski1997, pryde2003, beige2005}, quantum key distribution \cite{cerf2002}, and foundational studies of quantum mechanics \cite{cabello2010}. For a sense of the current state of the art, $16\times 16$ beam splitters are already commercially available and can be readily used in such integrated systems.

A qubit unitary transformation on optical modes transforms two input modes $(k,l)$ into two output modes $(k',l')$ according to
\begin{equation}
\begin{pmatrix}
k' \\ l'
\end{pmatrix}
=
\begin{pmatrix}
e^{i\phi}\sin\theta & e^{i\phi}\cos\theta \\
\cos\theta & -\sin\theta
\end{pmatrix}
\begin{pmatrix}
k \\ l
\end{pmatrix}
\end{equation}
and allows measurements to be done in any other basis. With a multiport device, unitary operations, and therefore arbitrary projective measurements, can also be realized in higher-dimensional quantum systems, providing a means of performing full state tomography with a quantum circuit.

\section{Naimark extension and SIC-POVMs}

A general quantum measurement is defined mathematically by a set of positive semidefinite operators $E_i$ that sum to identity, i.e., $\sum_i E_i = I$, and is called a positive operator-valued measurement or POVM. The probability that the outcome associated with $E_i$ occurs is given by the Born rule
\begin{equation}
p(i) = \trace{\rho E_i}
\end{equation}
where $\rho$ is the density operator describing the state of the measured system. In the special case that the measurement operators are orthogonal, that is, $E_iE_j = \delta_{ij} E_j$, the measurement is said to be projective.

There is a sense in which no generality is lost when considering only projective measurements since as a consequence of Naimark's theorem \cite{kraus1983}, any measurement on a quantum system can be achieved with a joint projective measurement on the system and an ancilla, provided the ancilla is large enough and initialized to a known pure quantum state \cite{peres1990}.

In the finite-dimensional case, there is a simple procedure for determining the projective measurement that corresponds to a POVM. Suppose we have a POVM $\{ E_i: i=1,2,...,n \} $ and each POVM operator $E_i$ is proportional to a rank-one projection onto the subspace spanned by the vector $e_i \in \mathcal{H}_m$. Given that $n > m$ and the POVM elements are not mutually orthogonal, construct the $m \times n$ matrix
\begin{equation}
V = \left( e_1 \ldots e_n \right)
\end{equation}
and search for an $ (n-m) \times n$ matrix $W$ such that
\begin{equation}
U =
\begin{pmatrix}
V \\ W
\end{pmatrix}
\end{equation}
is an $n \times n$ unitary matrix. In this case, the projectors associated with the subspaces spanned by the columns of $U$ form the desired projective measurement.

Here we are interested in a particular class of POVMs such that the $d^2$ POVM elements $E_i$ are subnormalized projectors such that
\begin{equation}
\trace{E_iE_j} = \frac{d\delta_{ij} + 1}{d^2(d+1)}.
\end{equation}
These measurements are often referred to as symmetric, informationally complete POVMs or SIC-POVMs. The term \emph{symmetric} is related to the constant trace overlap between distinct elements while \emph{informationally complete} means that the statistics of a SIC-POVM can be used for full quantum state tomography. Given a SIC-POVM $\{ E_i = \frac{1}{d}\Pi_i \}$ with probability $p(i)$ for every outcome $i$, the quantum state can be reconstructed via the rule \cite{adf2007}
\begin{equation}
\label{eq.sicrep}
\rho = \sum_{i=1}^{d^{2}} \left[(d+1)p(i) - \frac{1}{d} \right]\Pi_{i}.
\end{equation}
Using the Naimark extension, we describe optical experiments for realizing a qubit and a qutrit SIC-POVM with multiport devices.

\section{Multiport Qubit SIC-POVM}

\begin{figure}[t]
\centering
\includegraphics[scale = 0.30]{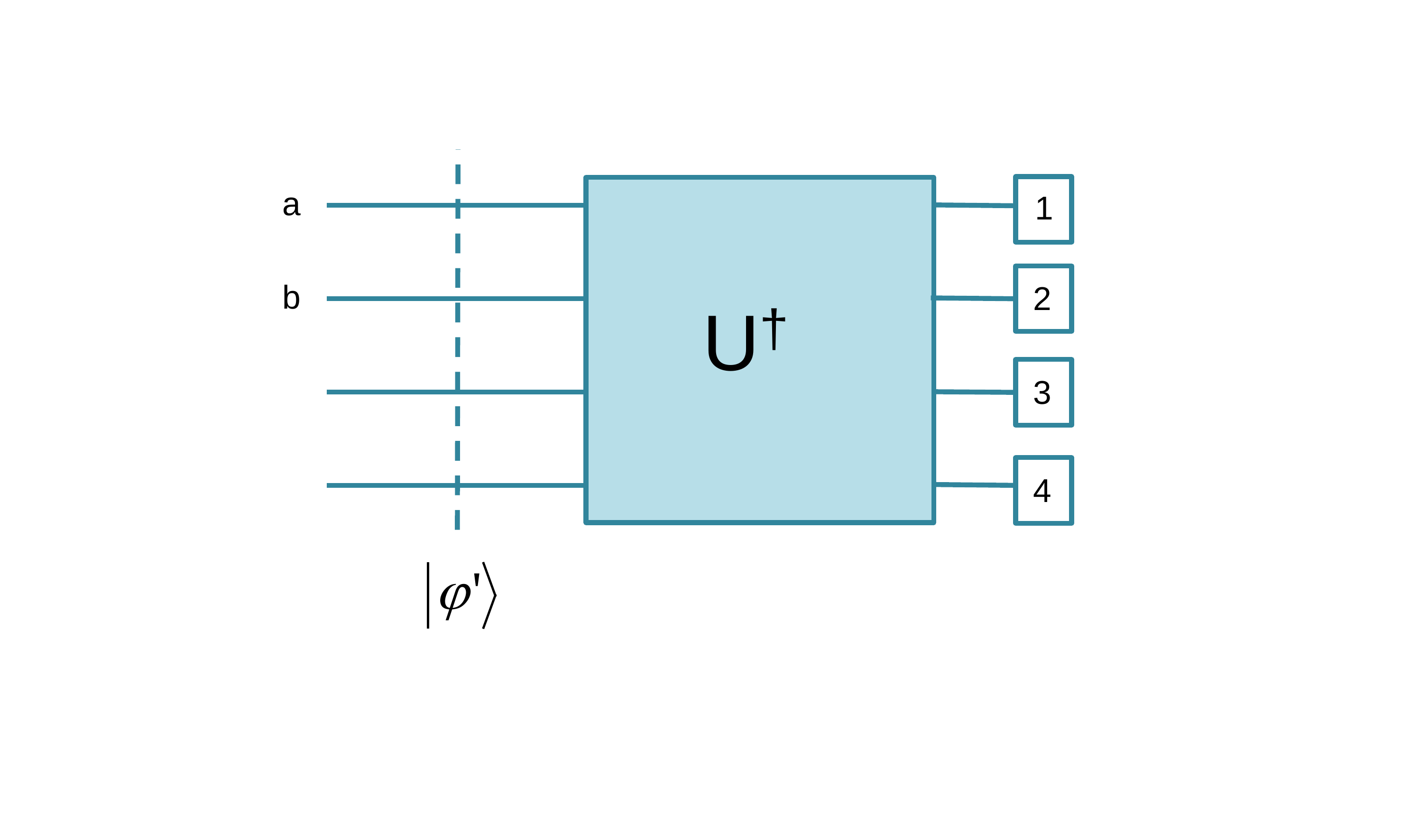}
\caption{Schematic for a multiport qubit SIC-POVM experiment. The unitary transformation $U^{\dag}$ is applied to the input state $\ket{\varphi'} = (a,b,0,0)^T$ and a projective measurement at the end yields statistics that correspond to the qubit SIC-POVM.}
\label{fig.qubitMultiExpt}
\end{figure}

A qubit SIC-POVM corresponds to a set of projection operators associated with the vertices of a regular tetrahedron circumscribed by the Bloch sphere. The choice for the SIC-POVM is not unique since one can always rotate the Bloch sphere but we may choose the projectors corresponding to the set of vectors
\begin{equation}
\label{eq.qubitSIC}
\left\{
\begin{pmatrix}
1 \\ 0
\end{pmatrix},
\frac{1}{\sqrt{3}}
\begin{pmatrix}
1 \\ \sqrt{2}
\end{pmatrix},
\frac{1}{\sqrt{3}}
\begin{pmatrix}
e^{i\frac{\pi}{3}} \\ e^{-i\frac{\pi}{3}}\sqrt{2}
\end{pmatrix},
\frac{1}{\sqrt{3}}
\begin{pmatrix}
e^{-i\frac{\pi}{3}} \\ e^{i\frac{\pi}{3}}\sqrt{2}
\end{pmatrix}
\right\}.
\end{equation}
This qubit SIC-POVM can be realized using an optical setup with path-encoded qubits, where information is encoded in the photon amplitude to traverse each mode.

Writing the qubit SIC-POVM vectors as a $2\times 4$ matrix, one particularly simple Naimark extension is given by the unitary matrix
\begin{equation}
U = \frac{1}{\sqrt{6}}
\begin{pmatrix}[1.25]
\sqrt{3} & 1 & e^{i \frac{\pi}{3}} & e^{-i\frac{\pi}{3}} \\
0 & \sqrt{2} & \sqrt{2}e^{-i\frac{\pi}{3}} & \sqrt{2}e^{i\frac{\pi}{3}} \\
\sqrt{3} & -1 & -e^{i\frac{\pi}{3}} & -e^{-i\frac{\pi}{3}} \\
0 & \sqrt{2} & -\sqrt{2} & -\sqrt{2}
\end{pmatrix}.
\end{equation}
The significance of $U$ for the multiport experiment is as follows. Denote the columns of $U$ by $\ket{U_{j}}$ and the vectors in Eq. (\ref{eq.qubitSIC}) by $\ket{u_j}$. If $\{ \ket{j}: j = 1, 2, \ldots, 4 \}$ is the standard basis in four dimensions then
\begin{equation}
U^\dag \ket{U_j} = \ket{j}.
\end{equation}
Now consider the path-encoded qubit $\ket{\varphi} = \left(a , b \right)^T$ depicted in Fig. \ref{fig.qubitMultiExpt}. Treating this qubit as the four-level state $\ket{\varphi'} = (a,b,0,0)^T$, it is straightforward to verify that
\begin{equation}
\left| \bracket{\varphi'}{U_j} \right|^2 = \frac{1}{2} \left|\bracket{\varphi}{u_j}\right|^2.
\end{equation}
This implies that the outcomes corresponding to the qubit SIC-POVM in Eq. (\ref{eq.qubitSIC}) are obtained by a projective measurement in the standard basis after applying $U^\dag$ to $\ket{\varphi'}$.

\begin{figure}[t]
\centering
\includegraphics[scale = 0.30]{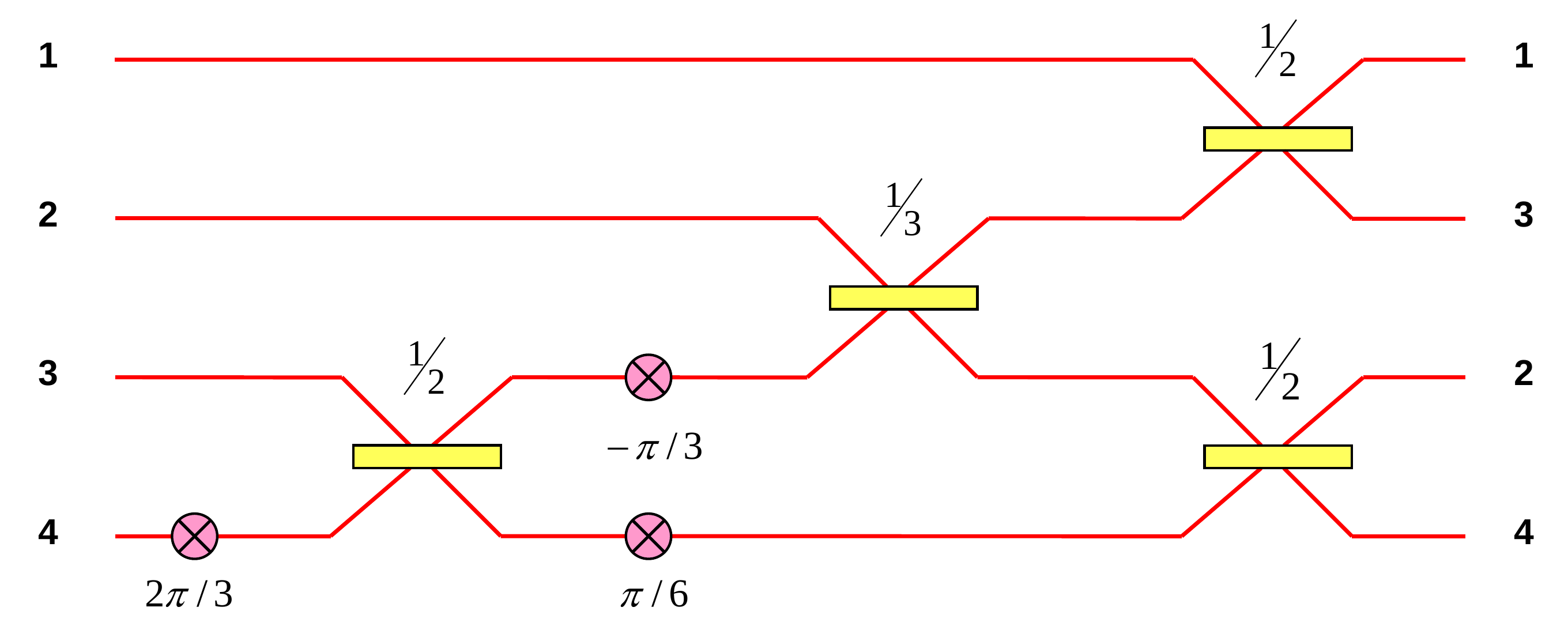}
\caption{Qubit SIC-POVM implemented with an optical multiport device. The linear optical network implements $U^\dag$ according to Eq.  (\ref{eq.decomposeU}). The labeling of the paths at the right indicates that a swap operator was applied, i.e., $U_3$.}
\label{fig.qubitOptics}
\end{figure}

In practice, $U^\dag$ can be implemented using a $4 \times 4$ multiport device. To derive an optical implementation for such a multiport, we need to decompose $U$ into a sequence of unitaries composed of $2\times 2$ blocks, each of which can then be performed by some combination of phase shifters and beam splitters on the appropriate optical modes. One such decomposition is achieved using the following matrices:
\begin{align}
U_1 &=  \frac{1}{\sqrt{2}}
\begin{pmatrix}
 1 & 0 & 1 & 0 \\
 0 & 1 & 0 & 1 \\
 1 & 0 & -1 & 0 \\
 0 & 1 & 0 & -1
\end{pmatrix}, \\
U_2 &= \frac{1}{\sqrt{3}}
\begin{pmatrix}
 \sqrt{3} & 0 & 0 & 0 \\
 0 & 1 & \sqrt{2} & 0 \\
 0 &- \sqrt{2} & 1 & 0 \\
 0 & 0 & 0 & \sqrt{3}
\end{pmatrix}, \\
U_3 &=
\begin{pmatrix}
 1 & 0 & 0 & 0 \\
 0 & 0 & 1 & 0 \\
 0 & 1 & 0 & 0 \\
 0 & 0 & 0 & 1
\end{pmatrix}, \\
U_4 &=
\begin{pmatrix}
 1 & 0 & 0 & 0 \\
 0 & 1 & 0 & 0 \\
 0 & 0 & e^{i \frac{\pi}{3}} & 0 \\
 0 & 0 & 0 &  e^{-i \frac{\pi}{6}}
\end{pmatrix}, \\
U_5 &=
\begin{pmatrix}
 1 & 0 & 0 & 0 \\
 0 & 1 & 0 & 0 \\
 0 & 0 & -1 & 0 \\
 0 & 0 & 0 & 1
\end{pmatrix}, \\
U_6 &= \frac{1}{\sqrt{2}}
\begin{pmatrix}
 \sqrt{2} & 0 & 0 & 0 \\
 0 & \sqrt{2} & 0 & 0 \\
 0 & 0 & 1 & 1 \\
 0 & 0 & 1 & -1
\end{pmatrix}, \\
U_7 &=
\begin{pmatrix}
 1 & 0 & 0 & 0 \\
 0 & 1 & 0 & 0 \\
 0 & 0 & 1 & 0 \\
 0 & 0 & 0 & e^{-i \frac{2\pi}{3}}
\end{pmatrix}.
\end{align}
It is straightforward to check that
\begin{equation}
\label{eq.decomposeU}
U = U_1 U_2 U_3 U_4 U_5 U_6 U_7.
\end{equation}
The optical multiport circuit that implements the qubit SIC-POVM according to Eq. (\ref{eq.decomposeU}) is shown in Fig. \ref{fig.qubitOptics}.

\section{Multiport Qutrit SIC-POVM}

\begin{figure}[t]
\centering
\includegraphics[scale = 0.27]{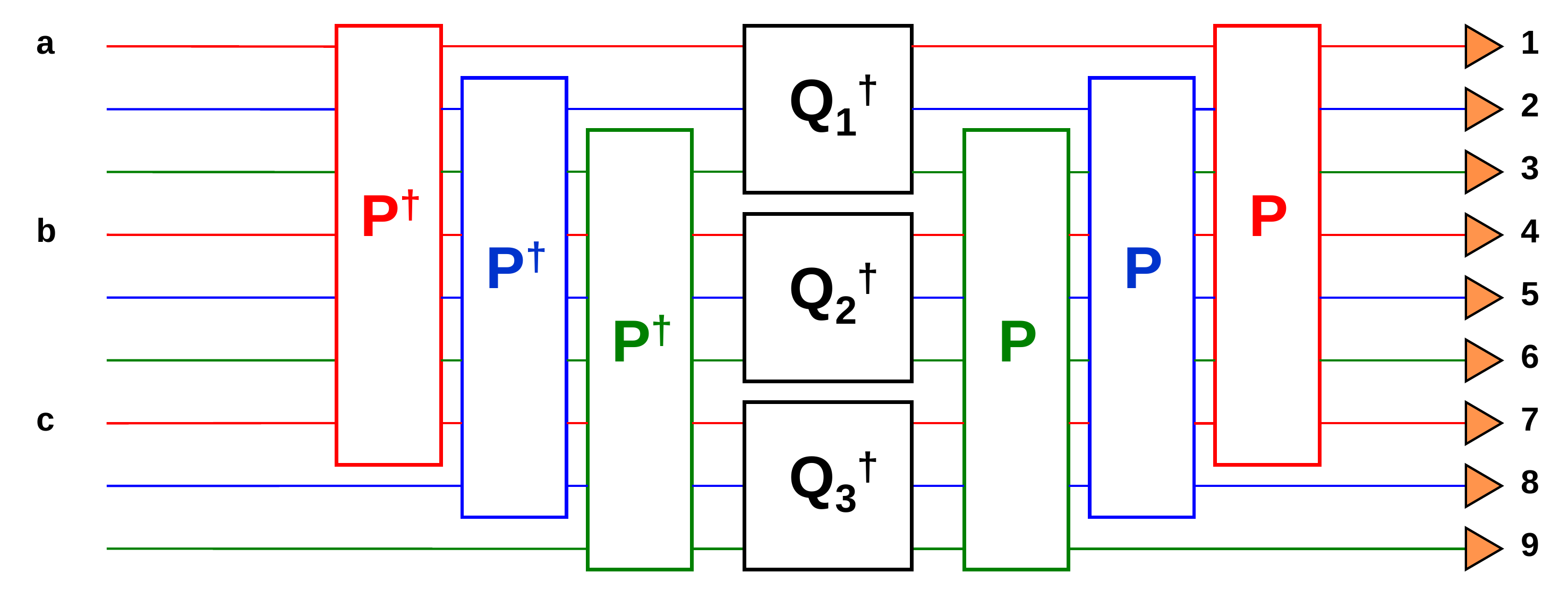}
\caption{(Color online) Schematic for a qutrit SIC-POVM generated by Eq. (\ref{eq.qutritSICfid}) using multiport devices. The input qutrit $(a,b,c)$ goes through modes 1, 4, and 7. The boxes $P$ and $P^\dag$ are performed on the modes as color coded in the diagram. The clicks in the numbered detectors correspond to SIC-POVM outcomes.}
\label{fig.qutritMultiExpt}
\end{figure}

Define the shift $X$ and phase $Z$ operators
\begin{equation}
X =
\begin{pmatrix}
0 & 1 & 0 \\
0 & 0 & 1 \\
1 & 0 & 0
\end{pmatrix},
\qquad
Z =
\begin{pmatrix}
1 & 0 & 0 \\
0 & \omega & 0 \\
0 & 0 & \omega^2
\end{pmatrix},
\end{equation}
where $\omega = e^{i \frac{2\pi}{3}}$. The qutrit SIC-POVM that we wish to implement is obtained by acting with $X^n Z^m$ for $n,m = 0,1,2$ on the vector
\begin{equation}
\label{eq.qutritSICfid}
\ket{\psi} = \frac{1}{\sqrt{2}}
\begin{pmatrix}
0 \\
1 \\
-1
\end{pmatrix}
\end{equation}
and taking the projection operators associated with the vectors generated.
The multiport experiment for implementing the SIC-POVM corresponding to $\ket{\psi}$ is shown in Fig. \ref{fig.qutritMultiExpt}.

To illustrate how the circuit works, consider an initial qutrit state $\ket{\phi} = \left(a,b,c\right)^T$. Similar to the qubit case, we think of $\ket{\phi}$ as the nine-level state $\ket{\phi'} = \left(a,0,0,b,0,0,c,0,0 \right)^T$. We want to perform a projective measurement on $\ket{\phi'}$ that gives the same statistics as the qutrit SIC-POVM corresponding to $\ket{\psi}$. In this case the measurement basis will be the columns of the following unitary matrix:
{\setlength{\arraycolsep}{1.75pt}
\begin{equation}
V = \frac{1}{\sqrt{6}}
\begin{pmatrix}
0 & 0 & 0 & -1 & -\omega ^2 & -\omega  & 1 & \omega  & \omega ^2 \\
\sqrt{2} & \sqrt{2} & \sqrt{2} & 0 & 0 & 0 & 0 & 0 & 0 \\
1  & \omega ^2 & \omega  & 1  & \omega  & \omega ^2 & 0 & 0 & 0 \\
1 & \omega  & \omega ^2 & 0 & 0 & 0 & -1 & -\omega ^2 & -\omega  \\
0 & 0 & 0 & \sqrt{2} & \sqrt{2} & \sqrt{2} & 0 & 0 & 0 \\
0 & 0 & 0 & 1 & \omega ^2 & \omega  & 1  & \omega  & \omega ^2 \\
-1 & -\omega ^2 & -\omega  & 1  & \omega  & \omega ^2 & 0 & 0 & 0 \\
0 & 0 & 0 & 0 & 0 & 0 & \sqrt{2} & \sqrt{2} & \sqrt{2} \\
1 & \omega  & \omega ^2 & 0 & 0 & 0 & 1 & \omega ^2 & \omega
\end{pmatrix}.
\end{equation}
}
The column vectors formed by taking elements in rows 1, 4, and 7 correspond to the qutrit SIC-POVM elements generated from $\ket{\psi}$. The significance of the unitary $V$ for the qutrit experiment is then completely analogous to that of $U$ in the qubit case. Denote the columns of $V$ by $\ket{V_{j}}$ and the normalized column vectors formed by taking rows 1, 4, and 7 of $V$ by $\ket{v_j}$. If $\{ \ket{j}: j = 1, \ldots, 9 \}$ is the standard basis in nine dimensions then
\begin{equation}
V^{\dag} \ket{V_j} = \ket{j}.
\end{equation}
Moreover, one can check that
\begin{equation}
\left| \bracket{\phi'}{V_j} \right|^2 = \frac{1}{3} \left|\bracket{\phi}{v_j}\right|^2
\end{equation}
so the outcomes of the standard basis measurement after applying $V^\dag$ yield statistics that is expected from the qutrit SIC-POVM.

The unitary $V^\dag$ can be realized with a $9 \times 9$ multiport device. However, the simplicity in the structure of $V$ allows us to perform the SIC-POVM using only $3 \times 3$ unitary operations. The decomposition of $V$ into a sequence of $3\times 3 $ unitaries is possible because $V$ exhibits the block-circulant structure
\begin{equation}
V =
\begin{pmatrix}
A & B & C \\
C & A & B \\
B & C & A
\end{pmatrix}
\end{equation}
that can be transformed into block diagonal form by the unitary matrix
{\setlength{\arraycolsep}{2pt}
\begin{equation}
S = P \otimes I_3
\equiv \frac{1}{\sqrt{3}}
\begin{pmatrix}
 1 & 0 & 0 & 1 & 0 & 0 & 1 & 0 & 0 \\
 0 & 1 & 0 & 0 & 1 & 0 & 0 & 1 & 0 \\
 0 & 0 & 1 & 0 & 0 & 1 & 0 & 0 & 1 \\
 \omega ^2 & 0 & 0 & \omega  & 0 & 0 & 1 & 0 & 0 \\
 0 & \omega ^2 & 0 & 0 & \omega  & 0 & 0 & 1 & 0 \\
 0 & 0 & \omega ^2 & 0 & 0 & \omega  & 0 & 0 & 1 \\
 \omega  & 0 & 0 & \omega ^2 & 0 & 0 & 1 & 0 & 0 \\
 0 & \omega  & 0 & 0 & \omega ^2 & 0 & 0 & 1 & 0 \\
 0 & 0 & \omega  & 0 & 0 & \omega ^2 & 0 & 0 & 1
\end{pmatrix}
\end{equation}}
where
\begin{equation}
\label{eq.unitaryP}
P = \frac{1}{\sqrt{3}}\left(
\begin{array}{ccc}
1 & 1 & 1 \\
\omega^{2} & \omega & 1 \\
\omega & \omega^{2} & 1
\end{array}
\right)
\end{equation}
is the three-dimensional Fourier matrix. The generalization of $P$ to all higher dimensions is known as the Bell multiport \cite{beige2005}, for which there is an efficient system for realizing the Fourier transform with a small number of optical elements \cite{zukowski1997}.

Thus,
\begin{equation}
\label{eq.unitaryQ}
Q = SVS^{\dag} \equiv  \left(
\begin{array}{ccc}
Q_{1} & 0 & 0 \\
0 & Q_{2} & 0 \\
0 & 0 & Q_{3}
\end{array}
\right)
\end{equation}
where
\begin{align}
Q_1 &= \frac{1}{\sqrt{6}}
\begin{pmatrix}
0 & \omega - \omega^{2} & \omega^{2} - \omega \\
\sqrt{2} & \sqrt{2} & \sqrt{2} \\
2 & -1 & -1
\end{pmatrix}, \\
Q_2 &= \frac{1}{\sqrt{6}}
\begin{pmatrix}
\omega^{2} - \omega & 0& \omega - \omega^{2} \\
\sqrt{2} & \sqrt{2} & \sqrt{2}  \\
-\omega^{2} & 2\omega^{2} & -\omega^{2}
\end{pmatrix}, \\
Q_3 &= \frac{1}{\sqrt{6}}
\begin{pmatrix}
\omega - \omega^{2} & \omega^{2} - \omega & 0 \\
\sqrt{2} & \sqrt{2} & \sqrt{2}  \\
 -\omega & -\omega & 2\omega
\end{pmatrix}.
\end{align}

Since $S$ and $Q$ are both composed of blocks of $3\times 3$ unitaries, this gives us the desired decomposition of $V$,
\begin{equation}
V = S^\dag Q S,
\end{equation}
where $Q$ has blocks acting on modes (1,2,3), (4,5,6), and (7,8,9), while $S$ has blocks acting on modes (1,4,7), (2,5,8), and (3,6,9) of the input state $\ket{\phi'}$.
The qutrit SIC-POVM requires $V^{\dag}$, which translates into the scheme displayed in Fig. \ref{fig.qutritMultiExpt}.
It is worth mentioning that because each $P$ acts on the output modes of a single tritter, it is more economical in practice to include them in the input state preparation, thereby reducing the depth of the circuit to just two parallel layers of optical elements. We also note that since all key components are $3 \times 3$ unitaries, a very compact realization of the qutrit SIC-POVM is possible using integrated $3 \times 3$ multimode interference devices.

Finally, we can construct an optical implementation for the multiport SIC-POVM by determining a quantum optical network for the unitaries $P^\dag$ and $Q^\dag_k,\ k = 1,2,3$. A linear optical network for $Q_k^\dag$ can be constructed from the Euler decomposition of a three-dimensional rotation, permutation matrices, and diagonal unitaries. We can write
\begin{equation}
\label{eq.decomposeQk}
Q_k^\dag = G_k R D_k,
\end{equation}
where $D_k$ are the diagonal matrices
\begin{align}
D_1 &=
\begin{pmatrix}
-i & 0 & 0 \\
0 & 1 & 0 \\
0 & 0 & 1
\end{pmatrix},
\\
D_2 &=
\begin{pmatrix}
-i & 0 & 0 \\
0 & 1 & 0 \\
0 & 0 & \omega
\end{pmatrix}, \\
D_3 &=
\begin{pmatrix}
-i & 0 & 0 \\
0 & 1 & 0 \\
0 & 0 & \omega^2
\end{pmatrix},
\end{align}
which can be performed using phase shifters alone; $R$ is the rotation matrix
\begin{equation}
R = \frac{1}{\sqrt{6}}
\begin{pmatrix}
0 & \sqrt{2} & 2 \\
\sqrt{3} & \sqrt{2} & -1 \\
-\sqrt{3} & \sqrt{2} & -1
\end{pmatrix};
\end{equation}
and $G_k$ are matrices that permutes the rows of $Q_k^\dag$,
\begin{align}
G_1 &=
\begin{pmatrix}
1 & 0 & 0 \\
0 & 1 & 0 \\
0 & 0 & 1
\end{pmatrix},
\\
G_2 &=
\begin{pmatrix}
 0 & 1 & 0 \\
 0 & 0 & 1 \\
 1 & 0 & 0
\end{pmatrix},
\\
G_3 &=
\begin{pmatrix}
 0 & 0 & 1 \\
 1 & 0 & 0 \\
 0 & 1 & 0
\end{pmatrix}.
\end{align}

\begin{figure}[t]
\centering
\includegraphics[scale = 0.35]{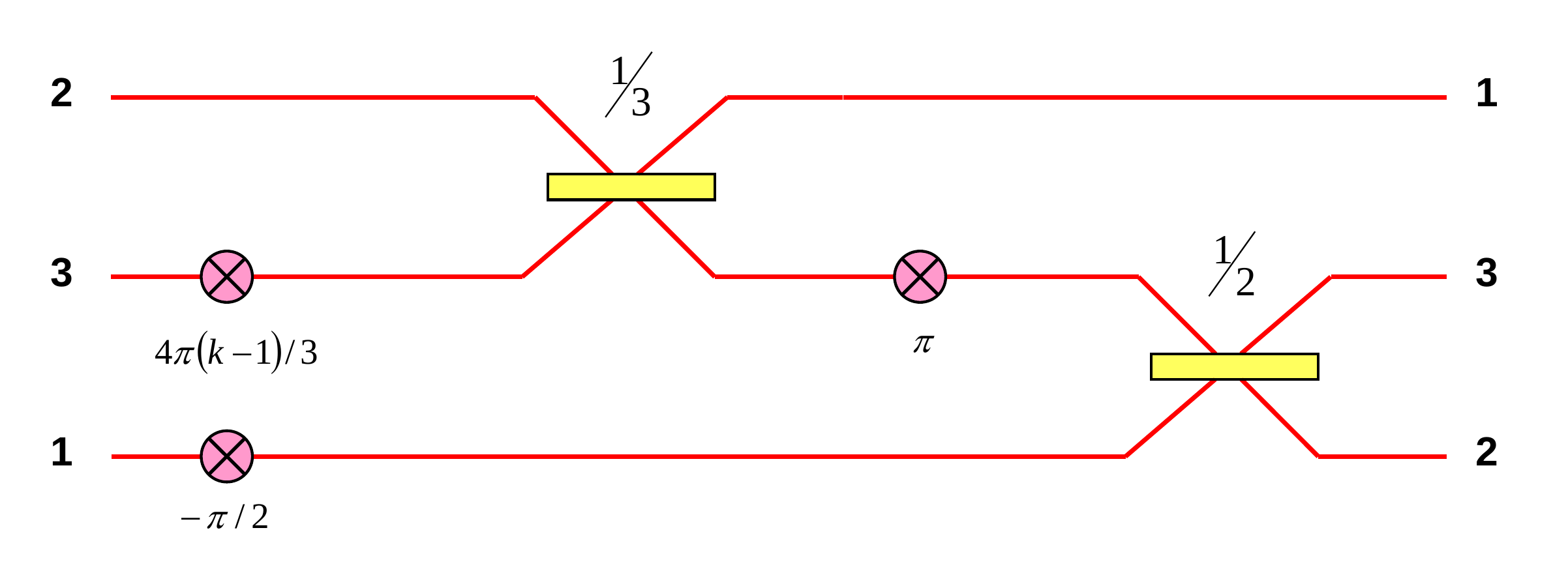}
\caption{\small Multiport circuit for the unitary operators $Q^\dag_k$. The circuits for $Q_k^\dag$ differ only by a phase shift and a relabeling of output modes as indicated by $G_k$. The output modes shown are for $Q_1^\dag$.}
\label{fig.qutritOptics}
\end{figure}

Using the Euler decomposition
\begin{equation}
\label{eq.eulerDec}
R = R_1(x) R_2(y) R_3(z)
\end{equation}
where
\begin{align}
R_1(x) &=
\begin{pmatrix}
 1 & 0 & 0 \\
 0 & \cos x & -\sin x \\
 0 & \sin x & \cos x
\end{pmatrix}, \\
R_2(y) &=
\begin{pmatrix}
\cos y & 0 & -\sin y \\
0 & 1 & 0 \\
\sin y & 0 & \cos y
\end{pmatrix}, \\
R_3(z) &=
\begin{pmatrix}
\cos z & -\sin z & 0 \\
\sin z & \cos z & 0 \\
0 & 0 & 1
\end{pmatrix},
\end{align}
the optical elements for our quantum circuit can be determined using the relation
\begin{equation}
\begin{pmatrix}
\cos\alpha & -\sin\alpha \\
\sin\alpha & \cos\alpha
\end{pmatrix}
=
\begin{pmatrix}
\sqrt{\epsilon} & \sqrt{1-\epsilon} \\
\sqrt{1-\epsilon} & -\sqrt{\epsilon}
\end{pmatrix}
\begin{pmatrix}
1 & 0 \\
0 & -1
\end{pmatrix}
,
\end{equation}
which shows how a two-dimensional rotation can be implemented using a beam splitter with reflectivity $\epsilon=\cos^2\alpha$ and a $\pi$-phase shifter. For the particular decomposition in Eq. (\ref{eq.eulerDec}), we have
\begin{equation}
x = -\frac{\pi}{4}, \ y = -\arccos\left(-\frac{1}{\sqrt{3}}\right),\ z = \frac{\pi}{2}.
\end{equation}

The unitaries $Q_k^\dag$ have almost identical optical circuits as prescribed by Eq.  (\ref{eq.decomposeQk}) since they differ only by a permutation and relative phase in one row. The quantum optical network for $Q_k^\dag$ is shown in Fig. \ref{fig.qutritOptics}. For example, with top-to-bottom path labeling, the circuit shown realizes $Q_1^\dag$. Only two phase shifters are needed because $R_3\left(\pi/2\right)$ is simply a swap operator sandwiched by two $\pi$-phase shifts acting on different paths, which collapses to a mere relabeling of paths.

The unitary $P$ transforms the modes into the Fourier basis and can be implemented using the same circuit as $Q_k^\dag$ since
\begin{equation}
P =
Q_1^\dag
\frac{1}{\sqrt{2}}
\begin{pmatrix}
1 & -1 & 0 \\
0 & 0 & \sqrt{2} \\
1 & 1 & 0
\end{pmatrix},
\end{equation}
that is, we just combine an equal beam splitter with $Q_1^\dag$.

\section{Practical quantum state tomography with a SIC-POVM}
\label{sec.pracEst}

\begin{figure}[t]
\centering
\includegraphics[scale=0.25]{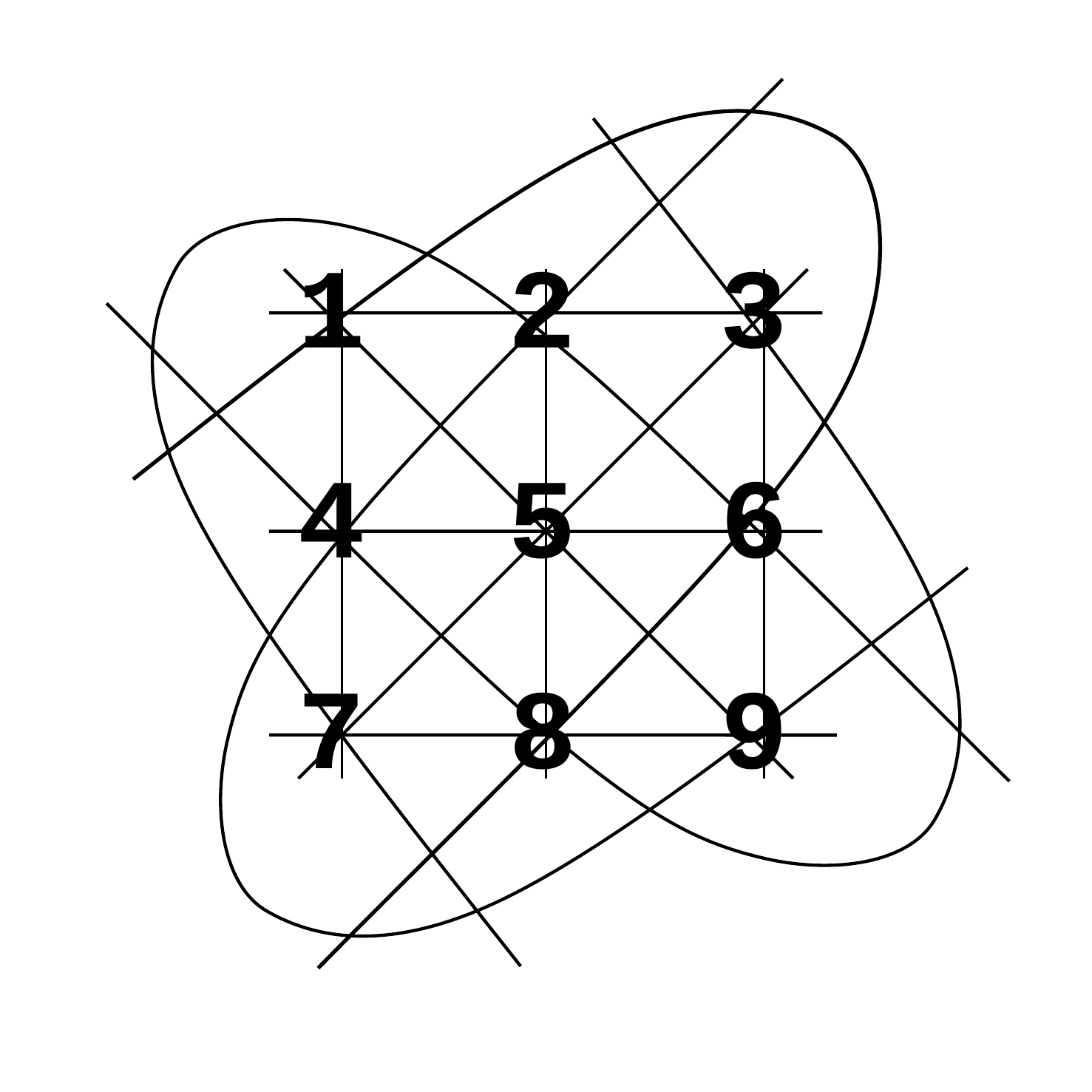}
\caption{\small \small The index triples $(ijk) \in Q$ for Eq. (\ref{eq.cubic}) depicted as twelve lines of an affine plane over $\mathrm{GF}(3)$.}
\label{fig.affineStriate}
\end{figure}

One standard application for the apparatus described above is for doing quantum state tomography---any procedure used for identifying the unknown state of a quantum system. In principle, an estimate of the quantum state is obtained from the measurement statistics of the qutrit SIC-POVM being performed on infinitely many copies of identically prepared systems. In practice, any physical measuring device is used a large but finite number of times and the frequencies $f(i)$ of each outcome are identified with the probabilities $p(i)$ expected in Eq. (\ref{eq.sicrep}). However, it is sometimes the case that the observed frequencies $f(i)$ deviate from the probabilities $p(i)$ in such a way that the associated operator $\rho$ is guaranteed to be a Hermitian operator but it is not necessarily positive semidefinite. Thus, one is required to introduce some modifications in the practical case of quantum state estimation from a finite number of measurements.

There is already a substantial body of work on reconstruction methods from observed frequencies \cite{paris2004, bk2010}. However, let us briefly consider a method adapted to the particular setting of SICs in the special case where the unknown quantum state is guaranteed to be a pure state. As such, the operator $\rho$ obtained from the relative frequencies must satisfy \cite{joneslinden2005}
\begin{equation}
\label{eq.pureState}
\trace{\rho^2} =
\trace{\rho^3} = 1.
\end{equation}

In the case of an unknown pure qutrit state estimated using the SIC-POVM (\ref{eq.qutritSICfid}), the conditions above imply two constraints on the manifold of allowed probability distributions. Substituting $\rho$ in Eq. (\ref{eq.sicrep}) into Eq. (\ref{eq.pureState}) and working out the algebra, we get the rather simple characterization
\begin{align}
\label{eq.quad}
\sum_i p(i)^2 &= \frac{1}{6}, \\
\label{eq.cubic}
\frac{1}{3} \sum_i p(i)^3 &= \sum_{(ijk)\in Q} p(i)p(j)p(k),
\end{align}
where $Q$ is the set of index triples $(ijk)$ obtained by taking the affine lines marked in Fig. \ref{fig.affineStriate}.

Any frequency distribution $f(i)$ obtained from the multiport SIC-POVM experiment that does not obey these two constraints for $p(i)$ can not properly describe a pure state. This suggests the practical procedure of choosing the probability vector $p(i)$ closest to $f(i)$ in the manifold of allowed probabilities defined by Eqs. (\ref{eq.quad}) and  (\ref{eq.cubic}) as our best estimate for the unknown pure state, e.g.,  with respect to Euclidean distance between $\vec{p}$ and $\vec{f}$,
\begin{equation}
\| \vec{p} - \vec{f} \| = \sqrt{\sum_i \left[p(i) - f(i) \right]^2}.
\end{equation}
The extension to mixed states is similar except that the manifold of valid probability distributions will be defined by a set of inequalities equivalent to imposing positivity on $\rho$, which is material for a future work \cite{tabia2012}.

\section{Discussion}

We have described an experimental scheme for directly implementing SIC-POVMs with integrated optics. It is worthwhile to comment on how the optical multiport scheme
serves as an improvement over previous techniques and what possible issues there are regarding its practical feasibility.

A key component for realizing the scheme is the SIC-POVM decomposition into optical networks. Designing the scheme for linear optics is practical given that various experimental groups are already currently working on multimode interference experiments with path encoded qudits but in principle the scheme can be realized with any setup that can perform the Naimark unitaries correspond to SIC-POVMs. As to the prescribed decomposition itself, we would like to emphasize that our decomposition shows a remarkable improvement over the general procedure for discrete unitary operators as described by Reck, et al. \cite{recketal1994}. Reck's method can be sketched in following way. Let $d$ be the Hilbert-space dimension and $N = d^2$. Then any $N$-dimensional unitary operator can be realized by an optical circuit with $(N-1)$ beam splitters between modes 1 and 2, $(N-2)$ beam splitters between modes 2 and 3, and so on up to one beam splitter between modes $(N-1)$ and $N$, a phase shifter accompanying each beamsplitter, and finally $(N-1)$ phase shifters before measuring the output. Counting all the optical elements needed, we obtain $(N^{2}-1)$ elements. Thus, by Reck's method no more than $(d^4-1)$ elements are required to realize the unitary for a $d$-dimensional SIC-POVM, but for particular unitary operators, it is possible that considerably fewer elements are needed. For example, Reck's method suggests that the $4\times 4$ unitary for the qubit SIC-POVM needs 15 elements, but our optical multiport scheme here uses only 7 elements. Similarly, the unitary corresponding to a qutrit SIC-POVM would generally require 80 elements but the one we present here uses only 44 elements.

We also remark on how the optical multiport scheme for qutrit SIC-POVMs compares to the Medendorp experiment \cite{medendorpetal2010}. In the Medendorp scheme, photons are sent into an optical storage loop and the SIC-POVM is performed by weakly coupled partially polarizing beam splitters (PPBS), with a full measurement completed after the signal makes 3 rounds. Because of the loop design, it uses only 22 elements for the measurement: 3 liquid crystal waveplates, 1 beam displacer, 3 PPBS, and the rest are half wave plates. Ideally, each PPBS extracts an infinitesimal portion of the signal that circles the loop infinitely many times. In the actual experiment, each PPBS reflected about 3\% of the horizontal polarization and 13\% of the vertical polarization. This led to a significant drop in visibilities even before we finished performing the SIC-POVM. Even with compensation wave plates that help tune the visibility of interference between modes, the visibilities achieved were 74\% and 36\% for the first and second rounds, respectively. Errors due to beam deformation and other losses in the loop were also corrected during the post-measurement analysis, but some of the results were still off by as much as 20\% from the theoretically expected values. An accurate assessment of the sensitivity of the multiport scheme to losses in the optical elements will depend on the actual apparatus used to realize it, but because there is no weak coupling demand, we expect it to be more robust to errors and show a vast improvement in performance compared to the Medendorp results.

To properly assess the feasibility of the multiport experiment, it is important to know how sensitive the scheme is to fabrication errors or losses in the optical elements. The specifics will depend on the particular apparatus used for the multiport scheme. For instance, one might employ commercially available micro-optical chips used in some telecom devices, which have built-in optical elements that then can be set to the appropriate values. Or, in more standard cases, one is likely to fabricate the integrated optics circuit himself. In any case, the biggest issue will be maintaining the interferometric phase stability among the optical modes. For example, in the qutrit case, the fact that $Q_{k}$ and $P$ act on different sets of modes means one has to actively check and correct for possibly unstable relative phases, which is quite a nontrivial task at present. A way to possibly avoid the phase stability issue is to fabricate the $9 \times 9$ unitary for the qutrit case as a single nine-mode integrated device, but since our decomposition involves 44 elements, that might create its own challenges.

\section{Summary and Outlook}

Integrated optics provides a highly promising platform for implementing general quantum circuits in an efficient and scalable manner. In particular, it offers the possibility of realizing complicated linear optical networks using multiport devices, allowing us to conduct experiments with higher-dimensional photonic systems \cite{recketal1994, polster2012, schaeff2011, withford2012, schaeffetal2012, spagnolo2012, laing2010, obrien2011}. Here we presented a practical design for performing SIC-POVMs on path-encoded qubits and qutrits using optical multiports. The prospect of being able to perform SIC-POVMs directly with multiports might open up new avenues of research in higher-dimensional quantum systems, an essential step in improving our understanding of quantum information and quantum foundations.

\begin{acknowledgments}
The author is grateful to Sven Ramelow and Christoph Schaeff for illuminating discussions on multiport devices. The author also thanks Chris Fuchs, Marcus Appleby, and \r{A}sa Ericsson for useful discussions and for reviewing drafts of the manuscript. The author is also grateful to the anonymous referee who pointed out a flaw in an earlier version of the qutrit optical circuit. This work was supported in part by the U.S. Office of Naval Research (Grant No.\ N00014-09-1-0247). Research at Perimeter Institute is supported by the Government of Canada through Industry Canada and by the Province of Ontario through the Ministry of Research and Innovation.
\end{acknowledgments}

\bibliography{qutritExptSIC}

\end{document}